\documentstyle[aps,multicol,pra,eqsecnum,epsfig]{revtex}
\begin{document}
\draft
\title{The ultimate quantum limit to the linewidth of lasers}
\author{H.M. Wiseman}
\address{Centre for Laser Science, Department of Physics, The 
University of Queensland, Queensland 
4072 Australia.}
\maketitle

\begin{abstract}
The standard quantum limit to the linewidth of a laser for which the 
gain medium can be adiabatically eliminated is $\ell_{0}=\kappa/2\bar{n}$. 
Here $\kappa$ is the intensity damping rate and $\bar{n}$ the mean 
photon number. This contains equal contributions from the loss 
and gain processes, so that simple arguments which attribute the 
linewidth wholly to phase noise from spontaneous gain are wrong. I 
show that an {\em unstimulated} gain process actually introduces no 
phase noise, so that the ultimate quantum limit to the laser 
linewidth comes from the loss alone and is equal to $\ell_{\rm ult}=
\kappa/4\bar{n}$.
I investigate a number of physical gain mechanisms which attempt to 
achieve gain without phase noise: a linear atom-field coupling with finite 
interaction time; a nonlinear atom-field coupling; and 
adiabatic photon transfer using a
counterintuitive pulse sequence. The first at best reaches the 
standard limit $\ell_{0}$, the second reaches $\frac34 \ell_{0}$, while 
the third reaches the ultimate limit of $\ell_{\rm ult} = \frac12 \ell_{0}$. 

\end{abstract}  

\newcommand{\beq}{\begin{equation}} 
\newcommand{\eeq}{\end{equation}}
\newcommand{\bqa}{\begin{eqnarray}} 
\newcommand{\eqa}{\end{eqnarray}}
\newcommand{\nn}{\nonumber} 
\newcommand{\nl}[1]{\nn \\ && {#1}\,}
\newcommand{\erf}[1]{Eq.~(\ref{#1})}
\newcommand{\dg}{^\dagger}
\newcommand{\rt}[1]{\sqrt{#1}\,}
\newcommand{\smallfrac}[2]{\mbox{$\frac{#1}{#2}$}}
\newcommand{\bra}[1]{\langle{#1}|} 
\newcommand{\ket}[1]{|{#1}\rangle}
\newcommand{\ip}[1]{\langle{#1}\rangle}
\newcommand{\sch}{Schr\"odinger } 
\newcommand{\schs}{Schr\"odinger's }
\newcommand{\hei}{Heisenberg } 
\newcommand{\heis}{Heisenberg's }
\newcommand{\half}{\smallfrac{1}{2}} 
\newcommand{\bl}{{\bigl(}}
\newcommand{\br}{{\bigr)}} 
\newcommand{\ito}{It\^o }
\newcommand{\str}{Stratonovich } 
\newcommand{\st}{Schawlow-Townes }

\begin{multicols}{2}

\section{Introduction}

It is more than 40 years since Schawlow and Townes introduced the 
idea of an ``optical maser'' \cite{SchTow58}, 
now known of course as a laser. Probably the most famous result from 
this paper is the expression for the quantum-limited laser linewidth, 
their Eq.~(17),
\beq
\Delta\omega_{\rm osc} = \frac{2\hbar\omega}{P_{\rm out}}(\Delta \omega)^{2}.
\eeq
Here $\Delta\omega_{\rm osc}$ is the half-width at half maximum (HWHM) of the 
laser, $\Delta \omega$ is the HWHM of the relevant atomic transition, 
$P_{\rm out}$ is the output power and $\omega$ the frequency of the laser. 
Defining $\ell = 2\Delta\omega_{\rm osc}$ 
and $\gamma = 2\Delta \omega$, this expression can be rewritten
\beq \label{STlb}
\ell_{\rm ST} = \frac{\hbar\omega}{P_{\rm out}}\gamma^{2},
\eeq
where ST stands for \st.
The derivation of this expression assumes reabsorption of photons by 
atoms in the ground state of the relevant transition is negligible, and also 
ignores thermal photons and other extraneous noise sources.

To accurately describe lasers a number of refinements must be made 
to the \st expression \cite{Hak84}. These are discussed in the Appendix. 
This discussion,
I believe, helps to put in perspective some of the past work 
on quantum limits to the laser linewidth. The end 
result is that a better expression for the standard quantum limit to 
the laser linewidth is  
\beq 
\ell_{\rm st} = \frac{\ell_{\rm bare}}{2\bar{N}} \leq \frac{\hbar\omega}{2P_{\rm 
out}}\frac{\gamma^{2}\kappa^{2}}{(\gamma+\kappa)^{2}}.
\eeq
Here st stands for standard (quantum limit). 
As explained in the Appendix, $\ell_{\rm bare}$ is the 
bare linewidth, $\bar{N}$ is the 
number of coherent excitations stored in the laser mode and its gain 
medium and $\kappa$ is the cavity linewidth. The inequality is an 
equality only for perfectly efficient output coupling.

In the limit $\gamma \gg \kappa$ the gain medium can be 
adiabatically eliminated, resulting in Markovian evolution for the 
laser mode. This means that $\bar{N}$ can be replaced 
by $\bar{n}$ (the mean photon number), and $\ell_{\rm bare}$ by 
$\kappa$, to give the standard Markovian limit as
\beq \label{ell0}
\ell_{0} = \frac{\kappa}{2\bar{n}}.
\eeq
For the remainder of this paper I will assume 
the Markovian limit, and drop the adjective ``Markovian'' distinguishing 
$\ell_{0}$ from $\ell_{\rm st}$ when no confusion is likely to arise. 

Most older textbooks \cite{Lou73,SarScuLam74,Lou83} quote the result 
in \erf{ell0}, 
or one which reduces to it in the appropriate limit of neither reabsorption 
nor thermal photons. The first two of these \cite{Lou73,SarScuLam74} 
derive this result 
rigorously using quantum Langevin equations. 
All three attempt to explain it in terms of 
the noise added by the spontaneous contribution to the (mostly 
stimulated) gain of photons from the atomic medium. Loudon 
\cite{Lou83} even 
recommends the argument based on the uncertainty 
principle given by Weichel \cite{Wei76}. 

The argument of Weichel is as 
follows (in my notation). In a laser at steady state, 
the rate of spontaneous emissions to total (spontaneous and 
stimulated) emissions is $1:\bar{n}+1$. Since the total gain rate must 
equal the total loss rate $\kappa \bar{n}$, 
the rate of spontaneous emissions is 
\beq
A = \frac{\kappa \bar{n}}{1+\bar{n}}  \simeq \kappa,
\eeq
where it is assumed that $\bar{n}\gg 1$. Now the reciprocal of this, 
$\Delta t = 1/\kappa$,  is \cite{Wei76} ``the average time between phase 
fluctuations caused by spontaneous emissions into the mode.'' 
Invoking 
the uncertainty principle
\beq
\Delta E \Delta t \geq \hbar/2,
\eeq
with the energy uncertainty of the mode being $\Delta E \simeq \bar{n} 
\hbar\Delta \omega_{\rm osc}$ gives
\beq
\ell = 2\Delta\omega_{\rm osc} = \frac{1}{\bar{n}\Delta t} = 
\frac{\kappa}{\bar{n}},
\eeq
which agrees with the Schawlow-Townes result (\ref{STlb}), with $\gamma$ 
replaced by $\kappa$ and $P_{\rm out}$ by $\kappa\bar{n}$.

The problem with all such simple arguments is that they put the blame 
for phase diffusion solely on the gain mechanism. This is an artifact 
of thinking in terms of normally-ordered operator products. 
That is, it results from using (implicitly in most cases) the 
Glauber-Sudarshan $P$ function \cite{Gla63,Sud63,WalMil94} 
as a true representation of the 
the fluctuations in the laser mode field. The $P$ function is of 
course no more fundamental than the $Q$ function \cite{WalMil94}, which is a 
representation based on anti-normally 
ordered statistics. If one were to use  
the $Q$ function as an aid to intuition, one would find that it is 
the loss process that is wholly responsible for the phase noise. Of 
course the rate of phase diffusion would agree with that from the $P$ 
function, at least in steady-state where loss and gain balance. 

If one asks a question about phase diffusion, the only objective 
answer will come from using the phase basis itself. This is far more 
difficult than using the more familiar phase-space representations, 
but some approximate results have been obtained \cite{BarStePeg89}. These show 
that, at steady state, the phase diffusion has equal contributions 
from the loss and gain process. The same result occurs from a Wigner 
function calculation \cite{BarStePeg89}. This is not surprising 
since symmetrically 
ordered moments are known to closely approximate the true moments 
for the phase operator for states with well-defined amplitude 
\cite{HilFreSch95}. 

The fact that phase diffusion comes equally from the loss and gain 
processes suggests that the standard quantum limit to the laser 
linewidth, $\ell_{0}$ of \erf{ell0}, may not be the ultimate quantum 
limit. The contribution from the loss mechanism is unavoidable. A 
laser, at least in useful definitions~\cite{Wis97}, requires  
linear damping of the laser mode in order to form an output beam.
However, it may be that the standard gain mechanism could be replaced 
by some other gain mechanism that causes less phase diffusion. The 
ultimate quantum limit to the laser linewidth could thus be as small 
as one half of the standard limit.

In this paper I investigate various gain mechanism in an attempt to 
find one which causes less phase diffusion than the standard gain 
mechanism. First, in Sec.~II, I review the standard (ideal) model for 
a laser, giving rise to the standard quantum limit $\ell_{0}$. Next, 
in Sec.~III, I present gain without stimulated emission, which 
produces a linewidth of $\ell_{0}/2$, and discuss how this 
can be physically realized. In Secs.~IV and V I present models which 
attempt to approximate gain without stimulated emission, using a 
micromaser-like interaction and a nonlinear field-atom interaction 
respectively, and discuss their success. 
After a comparison of these results in Sec.~VI, I conclude in 
Sec.~VII by returning to a derivation of the quantum limits to the 
laser linewidth using an uncertainty relation.

\section{Ideal standard laser}
    
\subsection{The laser model}\label{1b1}

The ideal standard laser master equation results from just about any 
gain medium under the appropriate (that is to say, ideal) 
conditions. Here I will present probably the simplest derivation, in 
which the gain is due to the coupling of the laser mode with a 
single transition in an atom. Ignoring the other levels in the atom, 
the interaction Hamiltonian is
\begin{equation} \label{linHam}
H = i \Omega( \sigma a^\dagger -  \sigma ^\dagger a),
\end{equation}
where $a$ is the annihilation operator for the cavity mode, $\sigma = 
|l\rangle \langle u|$ is the lowering operator for the atom, and 
$\Omega$ is the one-photon Rabi frequency. 

Let the interaction 
time $\tau$ be such that $\epsilon \sqrt{\bar{n}} \ll 1$, where 
$\epsilon=\Omega\tau$ and $\bar{n}$ is the 
mean intracavity photon number. Then the unitary operator $\exp(-iH\tau)$ 
acting on the initially factorized state $R = \rho \otimes 
|u\rangle\langle u|$ 
can be expanded to second order in $\epsilon$ to give the entangled state 
for the atom and field
\begin{eqnarray}
R &=& 
\rho \otimes |u\rangle\langle u|  + \epsilon \left( a^\dagger \rho \otimes 
|l\rangle\langle u| + {\rm H.c.} \right) \nonumber \\
&& + \, \epsilon^2 \left(
a^\dagger \rho a \otimes |l\rangle \langle l| - \smallfrac{1}{2}  
\{ a a^\dagger, \rho \} \otimes |u\rangle \langle u| \right).
\end{eqnarray}

Say there is a detector which detects which state the atom is 
in immediately after it has interacted with the field.
If the outgoing atom is detected in the upper state, then the conditioned 
state of the field (the norm of which represents the probability of this 
detection result) is, to first order in $\epsilon^{2}$,
\begin{eqnarray}
\tilde{\rho}_u &=& \bra{u}R\ket{u} \,
 =\, \left(1 - \epsilon^2{\cal A}[a^\dagger] \right) \rho  \nonumber \\
 &=& \exp(-\epsilon^2 
a a^\dagger /2)  \rho \exp(-\epsilon^2 a a^\dagger/2), \label{rhou}
\end{eqnarray}
where the  superoperator ${\cal A}[c]$ is defined for an arbitrary 
operator $c$ by
\begin{equation}
{\cal A}[c]\rho = \smallfrac{1}{2}\{ c^\dagger c, \rho\}.
\end{equation}
If the atom is detected in the lower state (which happens rarely), the 
state is
\begin{equation} \label{rhol}
\tilde{\rho}_l = \bra{l}R\ket{l} = \epsilon^2 {\cal J}[a^\dagger] \rho,
\end{equation}
where the superoperator ${\cal J}[c]$ is defined by
\begin{equation}
{\cal J}[c]\rho = c\rho c^\dagger.
\end{equation}

If this were all that there was to the model then the 
master equation would be found simply by averaging over the two results. 
Assuming that 
excited atoms enter the cavity as a Poisson process with 
rate $\Gamma \ll \Omega$, the result would be
\beq \label{me1}
\dot{\rho} = \Gamma \epsilon^{2} {\cal D}[a\dg]\rho + \kappa {\cal 
D}[a]\rho.
\eeq
Here I have included linear loss (allowing the laser output) at rate 
$\kappa$, and I am using the notation
\beq
{\cal D}[c] \equiv {\cal J}[c] - {\cal A}[c].
\eeq
As long as $\Gamma \epsilon^{2} < \kappa$, this master equation has a 
steady state. However it is not an appropriate steady state for the 
device to be considered a laser. As discussed in Ref.~\cite{Wis97},
it is necessary to have $\bar{n} \gg 1$ for the output of the 
device to be coherent (in a quantum statistical sense). But in this 
limit, the stationary state of the master equation (\ref{me1})
has a photon number uncertainty $\sigma(n) \sim \bar{n}$. This leads 
to enormous low-frequency ($\sim \kappa/\bar{n}$) 
fluctuations in the intensity of the output 
beam. This ruins the second-order coherence of the device, which is 
why it could not be considered a laser \cite{Wis97}.

The origin of the problem with \erf{me1} is stimulated emission. 
Since this is part of the acronym l.a.s.e.r. it might be thought that 
this is a good thing, but it actually leads directly to the enormous 
intensity fluctuations inherent in \erf{me1}. This is because 
stimulated emission implies that, for $\bar{n} \gg 1$, the intensity gain  
is proportional to the intensity. Thus if the intensity fluctuates 
above its mean value that fluctuation will be reinforced by an 
increase in the gain, and if it fluctuates below the mean the gain 
will correspondingly decrease. To avoid this, and hence obtain the 
output characteristic of a laser, one actually wants a photon gain 
which is independent of fluctuations in the intensity.

In a real laser, this independence of gain rate with 
intensity occurs automatically as $\bar{n}$ becomes very large 
because of {\em gain saturation}. This is not difficult to derive 
in the master equation approach \cite{Wis93a}. However the result can 
be obtained more quickly in  the context of the present model, 
 by making sure that the atom gives up exactly one 
quantum of energy to the field, regardless of the field state.
This is achieved by the following procedure. 
If the atom is detected in the lower state, 
then the field has 
gained a photon and the process can stop. If it is detected in the 
upper 
state, one must try again with the same 
atom (or, more realistically, another excited atom). This process 
continues until the atom is detected in the ground state. 
This process is shown in Fig.~1.

Say $K$ 
atoms are required before the $(K+1)$th is detected in the lower 
state. From Eqs.~(\ref{rhou}) and (\ref{rhol}), 
the unnormalized state matrix after the $(K+1)$th atom is 
\begin{equation}
\tilde{\rho}_K = \epsilon^2 {\cal J}[a^\dagger] \exp[-K\epsilon^2 a 
a^\dagger /2]  \rho \exp[-K\epsilon^2 a a^\dagger/2].
\end{equation}
The norm of this state matrix is equal to the probability that this 
many atoms are needed. Thus, the average density operator, given that 
an atom is finally detected in the ground state, is
\begin{equation} \rho' = \sum_{K=0}^{\infty} \tilde{\rho}_K. 
\label{sum}
\end{equation}

Using the fact that $\epsilon^{2}$ is small, the sum in \erf{sum} 
can be converted to an 
integral by setting $\beta=\epsilon^2 K$:
\begin{equation}
\rho' ={\cal J}[a^\dagger] \int_0^{\infty} \exp(-\beta a a^\dagger /2)
 \rho \exp(-\beta a a^\dagger/2) d\beta. \label{b4i}
\end{equation}
This can be formally evaluated \cite{Wis93a} as 
\begin{equation}
\rho' = {\cal J}[a^\dagger] {\cal A}[a^\dagger]^{-1} \rho.
\end{equation}
The superoperator ${\cal A}[a^\dagger]^{-1}$ is well-defined 
because $a a^\dagger$ is a strictly positive operator \cite{seeBraCav94}.

The action of the superoperator ${\cal J}[a^\dagger]{\cal A}[a^\dagger]^{-1}$ is 
to add a photon to the system irrespective of its initial state. That 
is to say, it shifts the photon number distribution upwards by one. If 
this addition of a photon is assumed to occur at Poisson-distributed 
times, with a rate $\Gamma \ll \Omega\epsilon\bar{n}$,
then a Markovian master equation for the field results. 
If one also includes  
linear damping at rate $\kappa$ as above, and lets the gain (the rate of photon 
addition) be $\Gamma = \kappa\mu$, then one gets
\begin{eqnarray} \label{lme}
\kappa^{-1}\dot{\rho} &=& \mu\left({\cal J}[a^\dagger]{\cal 
A}[a^\dagger]^{-1}-1\right)\rho + {\cal D}[a]\rho \nonumber \\
&=& \mu{\cal D}[a^\dagger]{\cal A}[a^\dagger]^{-1}\rho + {\cal D}[a]\rho. 
\label{551a}
\end{eqnarray}
From \erf{b4i} and the identity
\beq
1 = \int_{0}^{\infty} du \exp(-u a a^\dagger /2) a a\dg \exp(-u a a^\dagger /2)
\eeq
it is easy to see that the ideal laser 
master equation (\ref{lme}), first derived in 
Ref.~\cite{Wis93a}, is of the 
required Lindblad form \cite{Lin76}.

\subsection{Stationary State}

In the fock basis the laser master equation (\ref{lme}) is
\bqa
\dot{\rho}_{n,m} &=& \mu\left( \frac{2\sqrt{nm}}{n+m}\rho_{n-1,m-1}
- \rho_{n,m}\right) - \frac{n+m}{2}\rho_{n,m} 
\nl{+} \rt{(n+1)(m+1)}\rho_{n+1,m+1} .
\eqa
Here, as in the remainder of the paper, I have set $\kappa = 1$.
Clearly the stationary state will be of the form $
\rho_{n,m} = \delta_{n,m}P_{n}$. 
The equation of motion for $P_{n}$ is
\beq
\kappa^{-1}\dot{P}_{n} = \mu\left(P_{n-1}-P_{n}\right) + (n+1)P_{n+1}-nP_{n}.
\eeq
This has the stationary solution $P_{n} = e^{-\mu} \mu^{n}/n!$.
That is, the intracavity photon statistics are exactly Poissonian.

The stationary state matrix can therefore be written
\beq
\rho_{\rm ss} = \sum_{n} e^{-\mu}\frac{\mu^{n}}{n!}\ket{n}\bra{n}.
\eeq
Equivalently, it can be written
\beq \label{sscoh}
\rho_{\rm ss} = \int_{0}^{2\pi} d\phi \ket{\alpha e^{i\phi}}\bra{\alpha 
e^{i\phi}},
\eeq
where $|\alpha| = \sqrt{\mu}$ and $\ket{\alpha e^{i\phi}}$ is a 
coherent state of amplitude $\ket{\alpha e^{i\phi}}$. From either 
expression it is easy to verify that 
the mean number is $
{\rm Tr}[a\dg a \rho_{\rm ss}] = \mu$
and the mean amplitude $
{\rm Tr}[a \rho_{\rm ss}] = 0$.

\subsection{Calculating the linewidth}

There are many different ways of calculating the linewidth of a laser 
from its master equation. One way is to covert the master equation 
into a Fokker-Planck equation for a quasiprobability distribution 
function such as the $P$, $Q$ or $W$ function \cite{Gar91}.  
This is relatively straightforward for a master equation of the form 
(\ref{lme}), despite the apparent awkwardness of the inverse 
superoperator ${\cal A}[a\dg]^{-1}$ \cite{Wis93a}. However, for other master 
equations as I will consider later in this paper, the conversion is 
not so simple. Therefore I will adopt a method using the Fock basis. 
The method is essentially a more rigorous version of 
that used by Sargent,  Scully, and 
Lamb \cite{SarScuLam74}.

The linewidth $\ell$ of a laser I have taken to be the FWHM of the 
Power spectrum
\beq
P(\omega) \propto \int_{0}^{\infty} d\tau \, g^{(1)}(\tau)\cos\omega \tau ,
\eeq
where the normalized first-order coherence function is
\beq
g^{(1)}(\tau) = \ip{a\dg(t+\tau)a(t)}_{\rm ss}/\ip{a\dg a}_{\rm ss}.
\eeq
If one represents the master equation (\ref{lme}) as $\dot\rho = {\cal 
L}\rho$ then one can write
\beq \label{g1p}
g^{(1)}(\tau) = {\rm Tr}[a\dg e^{{\cal L}\tau}(a\rho_{\rm ss})]/\mu.
\eeq

Note above that the stationary state matrix  $\rho_{\rm ss}$ 
is a mixture of coherent states, as in \erf{sscoh}. Since 
$g^{(1)}(\tau)$ is invariant under a phase shift, \erf{sscoh} 
implies that in \erf{g1p} one can take $
\rho_{\rm ss} = \ket{\alpha}\bra{\alpha}$,
with $|\alpha|^{2} = \mu$. Then \erf{g1p} becomes
\beq
g^{(1)}(\tau) = {\rm Tr}[a\dg \alpha \rho(\tau)]/\mu,
\eeq
where $\rho(t)$ obeys the master equation (\ref{lme}) and 
\beq
\rho(0) = \ket{\alpha}\bra{\alpha}.
\eeq

If one defines
\beq
f_{n}(t) =  \sqrt{n}\rho_{n-1,n}(t)/\alpha^{*}
\eeq
then one can write
\beq \label{g1pp}
g^{(1)}(t) = \sum_{n} f_{n}(t).
\eeq
Clearly if one can determine the evolution of $f_{n}(t)$, one can find 
$g^{(1)}(t)$ and hence the linewidth of the laser. 
From \erf{lme} one finds
\beq
\dot{f}_{n} = \mu\frac{2n}{2n-1}f_{n-1} - \mu f_{n} + n f_{n+1}
- \frac{2n-1}{2}f_{n}.
\eeq
Defining
\beq \label{defrn}
r_{n}(t) = \frac{\mu f_{n}(t)}{nf_{n+1}(t)},
\eeq
one obtains
\beq \label{dfn1}
\dot{f}_{n} = \left[ \frac{2n(n-1)}{2n-1}r_{n-1} -\mu + 
\frac{\mu}{r_{n}} - \frac{2n-1}{2}\right] f_{n}.
\eeq

Now from the definition (\ref{defrn}), 
$r_{n}(0) \equiv 1$.
Assuming that this ratio remains unity, expand  \erf{dfn1} to 
leading order in $1/\mu$ to get
\beq \label{dfn2}
\dot{f}_{n} \approx -\frac{1}{4n} f_{n}.
\eeq
Solving this and substituting into \erf{defrn} gives, to leading order,
\beq
r_{n}(t) \approx \exp\left( - \frac{t}{4n^{2}}\right) \approx 1 - 
\frac{t}{4n^{2}},
\eeq
where the expansion to first order is valid for times much less than 
$\mu^{2}$. Since, as will be shown, the coherence time $\sim 2/\ell$ 
is of order 
$\mu$, it is quite safe to make this expansion even for times long 
compared to the coherence time.

Substituting this expression for $r_{n}(t)$ into \erf{dfn1} gives the 
more accurate expression
\beq \label{dfn3}
\dot{f}_{n} \approx  -\frac{1}{4n}\left[1 + \frac{n-\mu}{n^{2}}t\right] f_{n}.
\eeq
Since the initial condition is
\beq
f_{n}(0) = e^{-\mu}\frac{\mu^{n-1}}{(n-1)!},
\eeq
the only significant contribution to the sum (\ref{g1pp}) comes from 
$n$ such that $|n-\mu| \alt \sqrt{\mu}$. Also, as noted above, one can 
assume $t \alt n$. Then the correction term in \erf{dfn3} is of order 
$\mu^{-1/2}$ and can be ignored. One can thus return  to the expression 
\erf{dfn2}, which becomes (again ignoring corrections of order 
$\mu^{-1/2}$),
\beq
\dot{f}_{n} \approx  -\frac{1}{4\mu} f_{n}.
\eeq

The first order coherence function is thus 
\beq
g^{(1)}(\tau) = \exp(-\tau/4\mu),
\eeq
so that the coherence time is $4\mu$ (which is of order $\mu$ as 
promised). The Fourier transform of this expression is a Lorentzian 
with FWHM 
\beq
\ell = \frac{1}{2\mu}.
\eeq
This is the standard quantum limit $\ell_{0}$ of the linewidth for an 
ideal laser.

\section{Gain without stimulated emission}

Since the ``stimulated emission of radiation'' is part of the acronym 
for laser, it might be thought that stimulated emission is essential 
to produce a laser. While a typical 
laser does rely upon stimulated emission to ensure that it runs 
single-mode, the fact that the model of 
Section~\ref{1b1} adds photons one by one suggests that it is not 
strictly necessary. 
I will now show that stimulated emission is indeed {\em not} necessary 
for laser action, and in fact that eliminating stimulated emission 
eliminates the phase 
diffusion caused by the gain process.

Stimulated emission is a simple 
consequence of the linear coupling of the laser field to its source, 
as in Eq.~(\ref{linHam}). That is to say, the Hamiltonian 
(\ref{linHam}) is linear in the annihilation operator $a$ which, for 
classical fields, can be replaced by the $c$-number $\sqrt{n}e^{i\phi}$. 
Whenever a rate is calculated in quantum theory 
it depends on the square of the Hamiltonian. Hence the fundamental 
gain rate from a linear coupling will vary as $n$, which is the 
so-called stimulated emission or Bose-enhancement factor. A fully quantum 
calculation of course gives spontaneous emission as well, and hence a 
gain rate proportional to $n+1$.

Since stimulated emission can be traced to the presence of $a$ in the 
coupling Hamiltonian, the only way to totally remove it is to substitute for 
$a$ a different 
lowering operator, one whose classical analogue does not increase with $n$. That 
is to say, in Eq.~(\ref{linHam}), replace
\begin{equation}
	a = \sum_{n=1}^{\infty}\sqrt{n}\ket{n-1}\bra{n} 
\end{equation}
by the Susskind-Glogower \cite{SusGlo64} $e\equiv \widehat{e^{i\phi}}$ operator
\begin{equation}
e = (a a^{\dagger})^{-1/2} a= \sum_{n=1}^{\infty} \ket{n-1}\bra{n}.
\end{equation}
The new Hamiltonian would be extremely 
nonlinear if expressed as a power series in $a$ and $a^{\dagger}$, but it cannot be 
denied that it will not exhibit any stimulated emission. 

Replacing $a$ by $e$ in the Hamiltonian (\ref{linHam}) presents no 
problems in the rest of the derivation. Moreover, it is not even 
necessary to assume that $\epsilon = \Omega \tau$ is very small. 
Instead, the result is independent of $\epsilon$, due to the  
fact that $e e^{\dagger} = 1$. In particular, if one chooses 
$\epsilon = \pi/2$, the transformation effected on the field by one 
transit of the atom is semi-unitary:
\beq
\exp\left[ \frac{\pi}{2}(e\dg \sigma - \sigma\dg e)\right] 
\ket{u}\ket{\psi} = \ket{l}S\ket{\psi}.
\eeq
Here $\ket{\psi}$ is the state of the field and
\beq
S = e\dg = \sum_{n=0}^{\infty} \ket{n+1}\bra{n}.
\eeq
The operator $S$ is semi-unitary rather than unitary 
because $S\dg S=1$, but $S S\dg =1-\ket{0}\bra{0}$.

Surprisingly, this transformation can be achieved 
physically using only the usual electric-dipole coupling \cite{Par95}. 
The trick 
is to use a three-level $\Lambda$ atom and another, classical field 
\cite{fn1}. 
 Then, using 
a using a counter-intuitive pulse sequence, the atom is transferred 
from one lower state to the other, and one photon is created in the 
cavity field (with the energy lost from the classical field). 
 Like the gain process in Sec.~II, this adds 
precisely one photon to the field. The difference is that it does this 
without entangling the state of the field and the atom, and hence 
leaves the state of the field pure. 

Taking the rate of addition of photons to the field to be $\Gamma$ as 
before, in place of Eq.~(\ref{lme}) one obtains
\begin{equation}
    \dot\rho = 
    \mu{\cal D}[e^{\dagger}]\rho + {\cal D}[a]\rho
	\label{lmetilde}.
\end{equation}
In the fock basis this becomes
\bqa
\dot{\rho}_{m,n} &=& \mu (\rho_{n-1,m-1}-\rho_{m,n})  - (n+m)\rho_{n,m}/2
\nl{-} \rt{(n+1)(m+1)}\rho_{n+1,m+1}  .
\eqa
This yields exactly the same equation for the diagonal elements (the 
photon number populations). Hence 
the unstimulated master equation produces exactly the same 
photon number statistics as does the standard laser master equation 
(\ref{lme}). 

To calculate the linewidth, proceed as before. One finds the 
following equation for $f_{n}$, defined as in Sec.~II:
\bqa
\dot{f}_{n} &=& \mu\left(\rt{\frac{n}{n-1}}f_{n-1} - f_{n}\right) + n f_{n+1}
- \frac{2n-1}{2}f_{n} \\
&=& \left[ \rt{n(n-1)}r_{n-1} -\mu + 
\frac{\mu}{r_{n}} - \frac{2n-1}{2}\right] f_{n}.
\eqa
Assuming $r_{n} \approx 1$ yields, as above, the self-consistent solution 
\beq
\dot{f}_{n} \approx  -\frac{1}{8\mu} f_{n}.
\eeq

The first order coherence function is therefore
\beq
g^{(1)}(\tau) = \exp(-\tau/8\mu),
\eeq
so that the linewidth is
\beq
\ell = \frac{1}{4\mu}.
\eeq
This is half the standard quantum limit $\ell_{0}$ of \erf{ell0}.
As explained in the introduction, the standard quantum limit for the 
phase diffusion rate contains equal contributions from the gain and 
loss processes. The gain process considered in this section does 
not introduce any phase noise; the operator $e\dg$ is more or less the 
exponentiation of the phase operator and so increases the photon 
number without affecting the phase distribution at all. Thus the phase
diffusion in this model comes wholly from the loss process, and the 
rate is half the standard rate.
%

\section{Finite atom-field interaction time}

The preceding section showed that an interaction in which the atom 
 is sure to give up its quantum of energy to the field {\em from a 
 single pass} results in a linewidth a factor of two smaller than the 
 standard limit. It was noted there that this could be achieved 
 using an adiabatic passage, but this has yet to be done 
 experimentally. This suggests that it would be worth exploring other 
 ways to mimic the unstimulated gain process.
 
In this section I investigate one idea, based upon the gain mechanism of 
a micromaser \cite{FilJavMey86,Rai94}. This utilizes the same 
Jaynes-Cummings  coupling (\ref{linHam}) as in Sec.~II. The 
difference is that the scaled interaction time $\epsilon = \Omega\tau$ is not 
assumed to be small. This modifies the results of Sec.~II as follows.
The state of the field 
conditioned on the detection of an atom in the lower state is 
\cite{FilJavMey86}
\beq
\tilde\rho_{l} = {\cal J}_{l}\rho,
\eeq
where 
\beq
{\cal J}_{l} = {\cal J}\left[e\dg \sin \left(\epsilon \rt{a 
a\dg}\right)\right].
\eeq
The field state conditioned on an atom passing through and remaining in 
the upper state is \cite{FilJavMey86}
\beq
\tilde{\rho}_{u} = {\cal J}_{u} \rho,
\eeq
where
\beq
{\cal J}_{u} = {\cal J}\left[\cos \left(\epsilon \rt{a 
a\dg}\right)\right].
\eeq

For states having a 
photon distribution localized around $\bar{n}$, if $\epsilon$ is 
such that $\epsilon \sqrt{\bar{n}} \approx \pi/2$, 
then it would seem that the action of the above superoperators could be 
approximated by
\bqa
{\cal J}_{l} &\approx& {\cal J}\left[e\dg\right], \\
{\cal J}_{u} &\approx& 0 .
\eqa
That is, the atom would almost certainly come out in the lower state, 
having given up its quantum of energy to the field. This is the same 
situation as for the unstimulated 
gain as shown in Sec.~III. This is why a finite interaction time 
$\epsilon$ might be expected to lead to a linewidth below the 
standard limit.  

If atoms are injected at a Poissonian rate $\mu$ then the total master 
equation is the usual micromaser master equation
\beq
\dot\rho = \left\{
\mu\left( {\cal J}_{u}+{\cal J}_{l} - 1\right) + {\cal D}[a]
\right\} \rho.
\eeq
Here linear damping at rate unity has been included also. This master 
equation has very complicated dynamics. For some values of $\epsilon$ 
and $\mu$ the stationary state does not have a well-defined intensity. 
That is, it is not the case that $\sigma(n) \ll \bar{n}$. 
Hence the device is not necessarily a true laser in the sense of 
Ref.~\cite{Wis97}.

To ensure that the a well-defined photon number distribution is 
produced, the same technique as in Sec.~II can be used. That is, if an 
atom is detected still in the upper state it is sent through again 
until it is detected in the lower state. The resulting master equation 
is
\bqa
\dot\rho &=& \left\{
\mu{\cal J}_{l}\sum_{k=0}^{\infty}{\cal J}_{u}^{k}  + {\cal D}[a]
\right\} \rho \\
&=& \left\{
\mu{\cal J}_{l}\left(1-{\cal J}_{u}^{k} \right)^{-1} + {\cal D}[a]
\right\} \rho.
\eqa
In the photon number basis
\bqa
\dot{\rho}_{n,m} &=& \mu \frac{\sin(\epsilon\sqrt{n})\sin(\epsilon\sqrt{m})}
{1-\cos(\epsilon\sqrt{n})\cos(\epsilon\sqrt{m})}\rho_{n-1,m-1}
\nl{-} \mu \rho_{n,m} - (n+m)\rho_{n,m}/2 \nl
+ \rt{(n+1)(m+1)}\rho_{n+1,m+1} .
\eqa

To find the linewidth, one proceeds as before to get the following 
equation for $f_{n}$:
\bqa
\dot{f}_{n} &=& \mu 
\frac{\rt{n}\sin(\epsilon\sqrt{n-1})\sin(\epsilon\sqrt{n})}
{\rt{n-1}\left[1-\cos(\epsilon\sqrt{n-1})\cos(\epsilon\sqrt{n})\right]}f_{n-1} 
\nl{-}  \mu f_{n} + n f_{n+1}
- \frac{2n-1}{2}f_{n}.
\eqa
Using the parameter
\beq
\phi \equiv \epsilon \sqrt{\mu},
\eeq
 one can continue the analysis as before and find eventually
\beq
\dot{f}_{n} \approx - \frac{1}{8\mu}\left[1 + 
\frac{\mu^{2}\sin^{2}(\phi/\mu)}{\sin^{2}\phi}\right] f_{n}.
\eeq
That is, the linewidth of the laser is found to be
\beq \label{microell}
\ell = \frac{1}{4\mu}\left[1 + 
\left(\frac{\sin(\phi/\mu)}{(\sin\phi)/\mu}\right)^{2}\right]
\eeq

It is easy to verify that this expression has a global minimum
\beq
\ell = \lim_{\phi \to 0}\frac{1}{4\mu}\left[1 + 
\left(\frac{\sin(\phi/\mu)}{(\sin\phi)/\mu}\right)^{2}\right] = \frac{1}{2\mu}.
\eeq
The limit $\phi\to 0$ is the limit of short interaction times in 
which the original model of Sec.~II is recovered, and the 
original linewidth $\ell_{0}$ also.
That is, no linewidth narrowing is possible using a finite 
interaction time in preference to an infinitesimal interaction time, 
despite the fact that the former can deposit a photon in the cavity 
in a single pass of the atom with very high probability. 

This line-broadening 
is definitely not an artifact of the assumption that the atom is 
always put through again if it is detected still in its upper state; 
a similar result is obtained for the usual master micromaser equation 
with a single pass per atom \cite{Scu91}. The approach to calculating 
the linewidth used in 
Ref.~\cite{Scu91} was similar to the one used here. A more accurate estimation 
of the  linewidth for 
the usual micromaser  has to take into account the fact that the 
intensity is not always well defined \cite{McGSch97}. This yields some 
deviations from the simple theory of Ref.~\cite{Scu91}, but still 
never shows any line-narrowing.

The reason that no linewidth narrowing occurs can be seen from the 
method of calculation I have employed. What turns out to be crucial is 
not to try to mimic the two terms in the ideal unstimulated gain, 
namely
\beq
{\cal D}[e\dg] = {\cal J}[e\dg]-{\cal A}[e\dg] = {\cal J}[e\dg]-1,
\eeq
but rather to mimic the following ratio of matrix elements 
involving these two terms:
\beq
\frac{\bra{n-1}\left\{{\cal J}[e\dg]\ket{n}\bra{n+1}\right\}\ket{n}}
{\bra{n}\left\{{\cal A}[e\dg]\ket{n}\bra{n+1}\right\}\ket{n+1}} = 1.
\eeq
In the unstimulated case the ratio is unity, and the difference from unity in 
other cases is proportional to the contribution to the linewidth from 
the gain process.
For the standard ideal laser,
\beq
\frac{\bra{n-1}\left\{{\cal J}[a\dg]\ket{n}\bra{n+1}\right\}\ket{n}}
{\bra{n}\left\{{\cal A}[a\dg]\ket{n}\bra{n+1}\right\}\ket{n+1}} \approx 1 - 
\frac{1}{8n^{2}}.
\eeq
Multiplying the deviation from unity by the gain constant $\mu$ 
and replacing $n$ by the mean photon number
$\mu$ gives $1/8\mu$. This is the standard contribution to the 
linewidth from the gain.
For the micromaser,
\beq
\frac{\bra{n-1}\left\{{\cal J}_{l}\ket{n}\bra{n+1}\right\}\ket{n}}
{\bra{n}\left\{\left[1-{\cal J}_{u}\right]\ket{n}\bra{n+1}\right\}\ket{n+1}} \approx 1 - 
\frac{\sin^{2}(\phi/\mu)}{8\sin^{2}\phi},
\eeq
which again explains the result in \erf{microell}.

\section{Nonlinear atom-field interaction}

With now a better understanding of how to reduce the gain-induced 
phase diffusion, I turn to a second method for trying to mimic 
unstimulated gain. As noted in Sec.~III, the operator $e\dg$ would 
require an infinite series to be expressed in terms of powers of 
$a$ and $a\dg$. Any Hamiltonian containing infinite powers of the 
field is unphysical. However, nonlinear optical processes containing 
field powers greater than unity do occur. This suggests that it is 
worth considering the following approximation
\bqa
e\dg &=& a\dg (a a\dg)^{-1/2} \,=\,  a\dg \left[ \mu + (a a\dg 
-\mu)\right]^{-1/2} \\
&\approx& \frac{a\dg}{\sqrt{\mu}}\left(\frac{3}{2} - 
\frac{1}{2}\frac{aa\dg}{\mu}\right) .
\eqa
That is, I wish to consider a nonlinear Jaynes-Cummings 
Hamiltonian of the form
\beq \label{nonlinHam}
H = i \Omega\left[ \sigma a\dg (3- aa\dg/\mu) + (3 - aa\dg/\mu) a 
\sigma \dg\right],
\eeq
which I expect to be useful when the photon number is approximately 
$\mu$. 

Physically, this Hamiltonian means that there are two processes which 
can excite the atom. The first is the usual linear dipole coupling to 
the field. The second is a three-photon process whereby a photon is 
virtually absorbed and re-emitted before finally being absorbed by the 
atom. The Hamiltonian matrix element for the second process is much 
smaller (for $\mu \gg 1$), which is physically reasonable, 
and is of the opposite 
sign. It is very doubtful that such a Hamiltonian could be achieved 
simply using a two-level atom. However, it is possible that an effective 
Hamiltonian of this form could be achieved using a multilevel atom, 
and other fields. I will not further discuss the feasibility of 
producing this Hamiltonian, as my chief concerned is with the 
question of principle: how well can the nonlinear Hamiltonian 
(\ref{nonlinHam}) reproduce the results of the unstimulated laser?

Assuming, as in previous sections, that the atoms are initially in the 
upper state and that any atom which exits the cavity still in the 
upper state is put through again, one derives, following the method 
of Sec.~II, the following master equation for the cavity mode.
\beq
\dot{\rho} = \mu{\cal D}[a^\dagger(3- aa\dg/\mu) ]{\cal 
A}[a^\dagger(3- aa\dg/\mu)]^{-1}\rho + {\cal D}[a]\rho.
\eeq
This has the same Poissonian mixture of number states as in the 
standard laser, and is amenable to the same method of calculating the 
linewidth. The result is
\beq
\ell = \frac{3}{8\mu}.
\eeq
That is, the contribution from the gain is $1/8\mu$, which is half 
the standard result and half the contribution of $1/4\mu$ from the 
loss (which is of course unchanged). This result can again be 
understood from the ratio
\beq
\frac{\bra{n-1}\left\{{\cal J}[a^\dagger(3- aa\dg/\mu) ]
\ket{n}\bra{n+1}\right\}\ket{n}}
{\bra{n}\left\{{\cal A}[a^\dagger(3- 
aa\dg/\mu)]\ket{n}\bra{n+1}\right\}\ket{n+1}} 
\approx 1 - 
\frac{1}{16n^{2}}. 
\eeq

\section{Discussion}

The standard quantum limit to the laser linewidth 
is not the ultimate quantum limit, even for the Markovian 
case in which the gain 
medium is eliminated from the equations of motion of the laser mode. 
Hidden within the standard Markovian expression 
\beq
\ell_{0}= \frac{\kappa}{2\bar{n}},
\eeq
are equal 
contributions of $\kappa/4\bar{n}$ from the gain and loss mechanisms for the 
laser. The latter contribution is a fundamental limit because linear 
loss is necessary for a coherent output beam to form. However the former 
results from a particular (extremely reasonable) assumption about the 
gain mechanism for laser action, namely that it comes from a weak linear 
coupling between the field and the gain medium. 

These arguments suggest that a different sort of gain mechanism could produce a 
laser with a linewidth up to $50\%$ below the standard quantum 
limit. As I have shown above, this ultimate Markovian limit 
\beq
\ell_{\rm ult} = \frac{\kappa}{4\bar{n}}
\eeq
can be achieved with a gain 
mechanism in which stimulated emission into the laser mode is eliminated. 
This requires that the 
matrix element for the 
addition of a photon to the laser mode be independent of the 
number of photons in the mode. 
As discussed, this could be physically 
achieved with adiabatic transfer of photons from another field using 
a counter-intuitive pulse sequence. 

I also examined two other gain mechanisms with similarities to the 
non-stimulated gain, to see if they also produced linewidth 
narrowing. The first, using the usual Jaynes-Cummings Hamiltonian but 
with a finite interaction time (as in the micromaser) did not. The 
second, using a nonlinear Jaynes-Cummings Hamiltonian involving 
three-photon as well as one-photon processes, produced a linewidth of 
\beq
\ell = \frac{3\kappa}{8\bar{n}}.
\eeq 
That is, the phase diffusion due to the 
gain was reduced by $50\%$ from the standard limit, resulting in an 
overall reduction of $25\%$ in the linewidth. Presumably higher-order 
nonlinear optical processes could more closely approach the ultimate 
limit. However, the difficulty in 
producing such nonlinear optical processes, and the fact that even a 
third-order nonlinearity fails to reach the ultimate limit, suggests 
that the adiabatic transfer is a better experimental option for
probing the ultimate quantum limit to the laser linewidth.

The ultimate limit for the rate of phase diffusion 
attained by eliminating gain noise can also be 
obtained, for short times,  
by instead eliminating loss noise. This can be achieved 
by coupling the laser output into a squeezed vacuum rather than a 
normal vacuum \cite{Gea87,MarWal88}. This only works for short times 
because it requires a specific  
phase relation between the squeezed vacuum and the coherent field in 
the laser, which will not remain valid since the laser phase continues to 
diffuse. It was suggested in Ref.~\cite{Gea87} that it might be 
possible to produce the squeezed vacuum by driving the squeezing 
device with the laser itself. In this case the whole squeezing device should 
really be considered as part (an internal absorber, in fact) 
of the laser, so that $\bar{n}$ in the 
original laser cavity should no longer be used as a good measure of the 
total stored excitation. Similar comments could be made about the 
proposal of Ghosh and Agarwal \cite{GhoAga89}, who also misquote the 
expression for the standard quantum limit given in 
Ref.~\cite{SarScuLam74} by a factor of two as their Eq.~(18). I 
believe that a rigorous analysis of these proposals would reveal 
no reduction below the standard quantum limit.

\section{Conclusion}

In the introduction I reproduced a simple argument purporting to use 
the time-energy uncertainty principle to derive the standard laser 
linewidth as a consequence of phase diffusion due to the gain process. 
The results of this paper show that any such simple argument is 
untenable since the gain process contributes only half of the 
standard phase diffusion rate. To compensate for disposing of this 
simple argument, I will conclude this paper with a (not quite so 
simple) argument deriving the ultimate Markovian quantum limit $\ell_{\rm ult}$ 
from another uncertainty principle argument. 

Instead of the 
energy-time relation, which is of doubtful content, I will use the 
quadrature uncertainty relation 
\beq
V(X)V(Y)\geq 1,
\eeq
where $X/2$ and $Y/2$ are the real and imaginary components of the laser 
mode amplitude $a$. Clearly the vacuum state is rotationally symmetric 
with
\beq
V(X)=V(Y) = 1,
\eeq
and this holds also for a coherent state (which is the state the laser 
mode can be assumed to be in). 

Let the mean amplitude of the coherent state be real and positive
so that $\bar{X}/2=\sqrt{\bar{n}}$ and $\bar{Y}=0$. 
The phase variance is
\beq
V(\phi) = V\left({\rm artan}\frac{Y}{X}\right) \simeq 
\frac{V(Y)}{\bar{X}^{2}} = \frac{1}{4\bar{n}}
\eeq
for $\bar{n}\gg 1$. Now the effect of linear damping for an 
infinitesimal time $dt$ is to reduce the mean photon number of the 
coherent state from $\bar{n}$ to $\bar{n}(1-\kappa dt)$. Thus the 
change in the phase variance is
\beq
dV(\phi) = \frac{\kappa dt}{4\bar{n}}.
\eeq

A noiseless gain process will return the mean photon number to 
$\bar{n}$ without increasing the phase noise. Therefore the phase 
variance increases at least as
\beq
V(\phi) \sim \frac{\kappa t}{4\bar{n}}.
\eeq
The linewidth is defined from the two-time correlation function 
\beq
\ip{a\dg(t)a(0)} \sim \bar{n}\ip{e^{i\phi(t)}} \sim \bar{n} 
e^{-V(\phi)/2} \sim \bar{n} e^{-\kappa t/8\bar{n}} .
\eeq
The Fourier transform of this expression is a Lorentzian with a FWHM 
of 
\beq
\ell = \frac{\kappa}{4\bar{n}},
\eeq
which is the ultimate quantum limit to the laser linewidth, as claimed.

\acknowledgments

I would like to thank D. Pope for a critical reading of this 
paper. The work was undertaken with the support 
of the Australian Research Council.

\appendix
\section*{Refining the Scawlow-Townes limit.}

The \st expression 
\beq \label{STl}
\ell_{\rm ST} = \frac{\hbar\omega}{P_{\rm out}}\gamma^{2}.
\eeq
was derived in the days 
before good optical cavities, 
and hence implicitly assumes that the atomic linewidth $\gamma$ is 
much smaller than the (FWHM) cavity linewidth $\kappa$. With $\kappa 
\alt \gamma$, it is necessary to replace $\gamma$ by the bare 
linewidth of the laser $\ell_{\rm bare}$. This is the frequency spread 
the output would have if the pump were suddenly turned off and all of 
the energy allowed to escape. For a large class of lineshapes, it can 
be shown that a reasonable approximation to the bare linewidth 
including contributions 
from the atomic (or other gain) medium and the cavity is
\beq \label{defelbare}
\ell_{\rm bare}^{-1} = \gamma^{-1}+\kappa^{-1}.
\eeq
For instance, this expression agrees with that given  by Haken (p.~103 
of Ref.~cite{Hak84}) for the case  where 
$\kappa \agt \gamma$. In the other cases, where $\kappa \ll \gamma$, the $\gamma$ 
in the \st expression is simply replaced by $\kappa$ \cite{Hak84}, 
which also agrees with \erf{defelbare}. The corrected \st expression 
is thus
\beq
\ell_{\rm ST}' = \frac{\hbar\omega}{P_{\rm out}}\ell_{\rm bare}^{2} = 
\frac{\hbar\omega}{P_{\rm out}}\frac{\gamma^{2}\kappa^{2}}{(\gamma+\kappa)^{2}}.
\eeq

The second correction which must be made to the \st linewidth relates 
to its use of the output power. Say, for argument's sake, that 
one one has a laser with 
linewidth  given by the \st limit, with all of the power coming out of 
one mirror. Then say that the mirror is replaced by one of the same 
reflectance, but with larger internal absorption. Then the power loss 
per round trip is identical, so the laser dynamics remain the same and 
the linewidth would remain the same. But the power out would be 
reduced because the transmittance is reduced. Therefore the \st 
formula would now predict an increased linewidth, which does not 
occur. In other words, the actual new linewidth would be {\em less} than the 
quantum limit set by the \st formula. It is obviously inappropriate 
that a quantum limit can be surpassed by building a worse device. 

The resolution to this problem with the \st linewidth is to eliminate 
$P_{\rm out}$ from the expression by recognizing that
\beq
\frac{P_{\rm out}}{\ell_{\rm bare}}
\eeq
is an {\em upper} bound on the mean energy $\bar{E}$ stored as coherent 
excitations in the laser system. 
If all of the stored coherent excitation eventually makes 
it into the output beam of the laser then the bare linewidth 
$\ell_{\rm bare}$ is due wholly to the output coupling and $P_{\rm 
out} = \ell_{\rm bare}\bar{E}$. In general $P_{\rm out}$ is less than 
this.  Reducing the output coupling efficiency (as discussed in the 
preceding paragraph) will not affect $\bar{E}$ so it is the correct 
parameter to use, rather than $P_{\rm out}$. The doubly corrected \st 
limit is thus
\beq
\ell_{\rm ST}'' = \frac{\ell_{\rm bare}\hbar\omega}{\bar{E}}
\leq \frac{\hbar\omega}{P_{\rm 
out}}\frac{\gamma^{2}\kappa^{2}}{(\gamma+\kappa)^{2}},
\eeq
where the inequality becomes an equality only for perfectly efficient output 
coupling. 

It is convenient to define the number of quanta of coherent 
excitation, $\bar{N}=\bar{E}/\hbar\omega$. For the case 
$\kappa\gg \gamma$ the excitation stored in the gain medium is 
negligible and $\bar{N} = \bar{n}$, where the latter represents the 
mean {\em photon} number in the cavity. If the gain medium cannot be 
adiabatically eliminated then $\bar{N}$ must include the excitations 
stored coherently in the gain medium as well. If $\gamma \ll \kappa$, 
as in the original \st expression, these excitations in the gain 
medium will be the dominant ones.

The final correction which needs to be made to the \st linewidth 
is to insert a factor of 
$\frac12$. The \st limit without this factor is appropriate to a laser 
below threshold in which the complex amplitude of the field undergoes 
large slow fluctuations (for $\bar{N}\gg 1$ which is the limit in 
which the \st equation is valid). Above threshold, the laser intensity 
fluctuations are almost eliminated \cite{Hak84}, leaving only phase 
fluctuations. This increases the coherence time by a factor of two, 
so that the final corrected expression for the laser linewidth is
\beq \label{ellst}
\ell_{\rm st} = \frac{\ell_{\rm bare}}{2\bar{N}} \leq \frac{\hbar\omega}{2P_{\rm 
out}}\frac{\gamma^{2}\kappa^{2}}{(\gamma+\kappa)^{2}}.
\eeq
Here st stands for standard (quantum limit) as opposed to ST which 
stands for \st.

In the limit $\kappa \ll \gamma$, which is the usual 
limit for most lasers, and which allows the gain medium to be 
adiabatically eliminated from the field equations, one obtains
\beq \label{ell0a}
\ell_{0} = \frac{\kappa}{2\bar{n}} \leq 
\frac{\hbar\omega}{2P_{\rm out}}\kappa^{2}.
\eeq
This result has often (including by myself \cite{Wis97}) been
quoted as the Schawlow-Townes limit, despite the 
obvious differences from \erf{STl}. Here I will call it instead the {\em 
standard Markovian quantum limit} to the laser linewidth.  
``Markovian'' refers to the fact that the equations of motion for the 
laser mode, including gain and loss, are well-approximated by 
Markovian equations. For the gain process this is a consequence of 
adiabatically eliminating the gain medium. For the loss process, it 
is simply a consequence of assuming a high-$Q$ cavity. 
Corrections (upwards) for non-Markovian loss (low-$Q$ cavities) 
are discussed for example in Ref.~\cite{PraAbb88}, 
but here I will always assume a high-$Q$ cavity. 

Obviously for 
$\gamma \alt \kappa$, the linewidth of \erf{ellst} will be less than 
the standard Markovian quantum limit of \erf{ell0a}. 
That is a reflection of the fact that in this case 
the bare linewidth $\ell_{\rm 
bare}$ is less than $\kappa$, and also that the gain medium 
is an extra reservoir of energy (coherent with 
the laser mode) so that $\bar{N}$ is greater than $\bar{n}$.
A linewidth which, in the ideal limit, reduces to $\ell_{\rm st}$ 
was recently derived in Ref.~\cite{ExtKupWoe95}, for a 
laser with $\gamma \alt \kappa$. These authors claimed that this was 
``reduced compared to the \st limit'' because they followed the 
common (but, in my opinion, erroneous) practice of  
identifying  $\ell_{0}$ as the
Schawlow-Townes limit. To me this seems to be an example of 
imprecise terminology obscuring an otherwise valuable contribution 
to fundamental laser physics. 
In this paper I always work with models in which 
the gain medium can be adiabatically eliminated, so that 
$\ell_{0}=\ell_{\rm st}$.

\end{multicols}

\newpage

\onecolumn

\begin{figure}[htbp]
\epsfig{figure=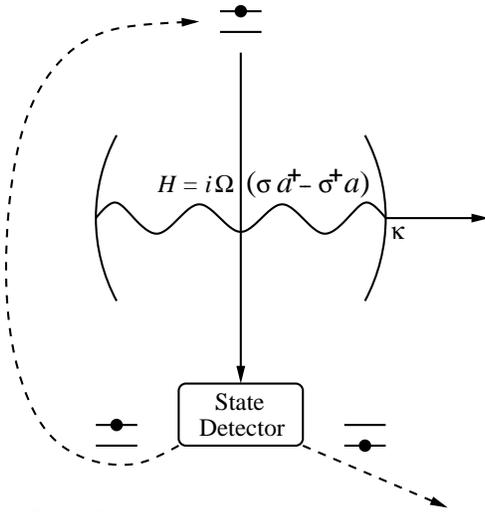,width=65mm}
\caption
{\narrowtext Schematic of an ideal laser. An atom in the upper state 
passes through the cavity and its state is then detected. 
If the atom remains in 
the upper state, the process is repeated until it is detected in the 
lower state. The time for this process (including repetitions)
is  assumed to be very short compared to the cavity damping time 
$\kappa^{-1}$. Once the atom is detected in the lower state, a new 
upper-state atom is injected after a random waiting time $\tau$ 
having an exponential distribution $w(\tau)=\exp(-\kappa\mu\tau)$. 
Here $\mu$ is the desired mean number of photons in the cavity. }
\end{figure}


\begin{references}
\bibitem{fn1}
It might be thought the presence of an external classical field would disqualify 
this device from the class of lasers, as defined in Ref.~\cite{Wis97}. 
However, it is in fact not 
necessary for the second field to have a well-defined coherent 
amplitude, or even a very large photon number, 
but only for it to have a negligible probability of 
being in the vacuum state.
\end{references}
\end{document}